\documentclass[
showpacs,
floatfix,
aps,  
prl,
twocolumn,
superscriptaddress,
amssymb,
]{revtex4-1}

\usepackage{times}
\usepackage{wasysym}
\usepackage{amssymb}
\usepackage{latexsym}
\usepackage[dvips]{graphicx}
\usepackage{graphicx}
\usepackage{dcolumn}
\usepackage{amsfonts}
\usepackage{bm}
\usepackage{epsfig}

\newcommand{\be}{\begin{equation}}
\newcommand{\ee}{\end{equation}}
\newcommand{\bea}{\begin{eqnarray}}
\newcommand{\eea}{\end{eqnarray}}

\newcommand{\pc}{{\mathcal P}}

\def\Eac{\mathcal{E}_{\mathrm{ac}}}

\def\m{m^\star}
\def\eac{\epsilon}

\def\psm{\varphi_{\it sm}}
\def\psh{\varphi_{\it sh}}

\def\epseff{\epsilon_{\mathrm{eff}}}

\def\oc{\omega_{\mbox{\scriptsize {c}}}}

\def\tq{\tau_{\rm q}}
\def\ttr{\tau}
\def\tem{\tau_{\rm em}}
\def\tsh{\tau_{\rm sh}}
\def\tsm{\tau_{\rm sm}}
\def\tee{\tau_{\rm ee}}
\def\tin{\tau_{\rm in}}
\def\tqim{\tau_{\mbox{\scriptsize {q0}}}}

\def\tst{\tau_\star}

\def\betao{\beta_{\omega}}

\def\mb{{m^\star_{\rm b}}}

\def\kex{\kappa}
\def\ksm{\kappa_{\rm sm}}
\def\ksh{\kappa_{\rm sh}}

\def\pex{\varphi}
\def\psm{\varphi_{\rm sm}}
\def\psh{\varphi_{\rm sh}}

\def\A{\mathcal{A}}

\newcommand{\req}[1]{Eq.\,(\ref{#1})}

\newcommand{\rfig}[1]{Fig.\,\ref{#1}}

\newcommand{\rref}[1]{Ref.\,\onlinecite{#1}}

\def\ma{0.0652}
\def\na{1.41}

\def\tqa{19.5}

\def\mb{0.0638}
\def\nb{2.15}

\def\tqb{20.0}

\def\mc{0.0631}
\def\nc{2.87}

\def\tqc{21.2}

\def\ne{n_{\rm e}}

\begin{document}
\title{
Effect of density on microwave-induced resistance oscillations in back-gated GaAs quantum wells 
}
\author{X. Fu}
\affiliation{School of Physics and Astronomy, University of Minnesota, Minneapolis, Minnesota 55455, USA}
\author{M.~D.~Borisov}
\affiliation{School of Physics and Astronomy, University of Minnesota, Minneapolis, Minnesota 55455, USA}
\author{M.~A.~Zudov}
\email[Corresponding author: ]{zudov001@umn.edu}
\affiliation{School of Physics and Astronomy, University of Minnesota, Minneapolis, Minnesota 55455, USA}
\author{Q.~Qian}
\affiliation{Department of Physics and Astronomy and Birck Nanotechnology Center, Purdue University, West Lafayette, Indiana 47907, USA}
\author{J.\,D. Watson$^\#$}
%\email[Present address: ]{Microsoft Station-Q at Delft University of // Technology, 2600 GA Delft, The Netherlands}
\affiliation{Department of Physics and Astronomy and Birck Nanotechnology Center, Purdue University, West Lafayette, Indiana 47907, USA}
\author{M.\,J. Manfra}
\affiliation{Department of Physics and Astronomy and Birck Nanotechnology Center, Purdue University, West Lafayette, Indiana 47907, USA}
\affiliation{Station Q Purdue, Purdue University, West Lafayette, Indiana 47907, USA}
\affiliation{School of Materials Engineering and School of Electrical and Computer Engineering, Purdue University, West Lafayette, Indiana 47907, USA}

\begin{abstract}
We report on microwave-induced resistance oscillations (MIROs) in a tunable-density 30-nm-wide GaAs/AlGaAs quantum well.
We find that the MIRO amplitude increases dramatically with carrier density.
Our analysis shows that the anticipated increase in the effective microwave power and quantum lifetime with density is \emph{not} sufficient to explain the observed growth of the amplitude.
We further observe that the fundamental oscillation extrema move towards cyclotron resonance with increasing density, which also contradicts theoretical predictions.
These unexpected findings reveal that the density dependence is not properly captured by existing theories, calling for further studies.
\end{abstract}
%\pacs{73.43.Qt, 73.63.Hs, 73.40.-c}
\received{7 November 2017; revised manuscript received 9 August 2018; published 17 September 2018}
\maketitle

%\section{Introduction}
Microwave-induced resistance oscillations (MIROs) appear in a two-dimensional (2D) electron gas (2DEG) \citep{zudov:2001a,ye:2001,karcher:2016} (or a 2D hole gas \citep{zudov:2014,shi:2014b}) subjected to low temperature $T$, weak magnetic field $B$, and microwave radiation of frequency $f =\omega/2\pi$.
It is well established experimentally that MIROs originate from the bulk of the 2DEG \citep{zhang:2007c,hatke:2008a,hatke:2008b,bykov:2010d,khodas:2010,fedorych:2010,andreev:2011,levin:2015,dorozhkin:2016b,herrmann:2016,herrmann:2017}.
Theoretically, microwave photoresistance $\delta R$ due to MIROs can be described by \citep{dmitriev:2005,dmitriev:2009b,dmitriev:2012}
\begin{equation}
\frac {\delta R (\eac)} {R_0} = -2\pi\eac\lambda^{2}\pc\eta \sin2\pi\eac\,.
\label{eq.miro}
\end{equation}
Here, $R_0$ is the resistance at $B = 0$, $\eac=\omega/\oc$, $\oc = e B/\m$ is the cyclotron frequency of the charge carrier with the effective mass $\m$, $\lambda = \exp (-\eac/2f\tq)$ is the Dingle factor, $\tq$ is the quantum lifetime, $\pc$ is the effective microwave power, and $\eta$ is the dimensionless parameter (discussed later in detail) which depends on the disorder characteristics and the inelastic relaxation.
The above expression was obtained assuming $2\pi k_B T  \gg \hbar \omega$ and is accurate away from the cyclotron resonance ($2\pi\eac \gg 1$), when the microwave power is not too high ($\pc \ll 1$), and when Landau levels are overlapping ($\lambda \ll 1$).  

To date, MIROs have been observed in three kinds of material systems, namely, GaAs/AlGaAs \citep{zudov:2001a,ye:2001}, Ge/SiGe \citep{zudov:2014,shi:2014b}, and MgZnO/ZnO \citep{karcher:2016} heterostructures, and the dependence on $\eac$, $\delta \rho_\omega \propto -\eac\lambda^2\sin2\pi\eac$, has been verified in many experiments. 
In addition, it was established that while MIROs at high microwave power significantly deviate from \req{eq.miro}, they can still be well described within the same theoretical framework after generalization to an arbitrary radiation intensity \citep{khodas:2010,hatke:2011e,shi:2017a}.
At the same time, experiments also revealed situations when existing theory is inadequate, e.g., in describing the measured dependencies on radiation polarization \citep{smet:2005,herrmann:2016} and on temperature \citep{shi:2016a}.
Limitations of the theory were also identified in the regime of separated Landau levels \citep{hatke:2011f} and in the radiation-induced modification of Shubnikov$-$de Haas oscillations \citep{shi:2015b}.

One important parameter, whose role has remained largely unexplored, is the carrier density $\ne$.
While it has been recently demonstrated that $\ne$ affects $\eac$, presumably through interaction-induced renormalization of the effective mass $\m$ \citep{shchepetilnikov:2017,fu:2017}, it should also modify other quantities, e.g., $\pc$ and $\eta$, entering \req{eq.miro}. 
Since the density dependencies of $\pc$ and $\eta$ are both known theoretically, MIRO measurements as a function of $\ne$ should provide an important test to existing microscopic description of microwave photoresistance.

%In this RC
In this Rapid Communication we investigate the effect of the carrier density on the MIRO amplitude employing a tunable-density 2DEG \citep{watson:2015,shi:2017c,fu:2017,qian:2017b}.
We find that the quantum lifetime depends on the carrier density only weakly, in agreement with the recent study investigating Shubnikov$-$de Haas oscillations in a similar device \citep{qian:2017b}.
Our main finding, however, is a significant \emph{growth} of the MIRO amplitude with the carrier density.
The analysis shows that this growth cannot be accounted for by the anticipated density dependence of $\pc\eta$ entering the prefactor of \req{eq.miro} \citep{khodas:2008,zhang:2014,dmitriev:2005,dmitriev:2009b,dmitriev:2012}.
Furthermore, we find that the fundamental extrema move towards $\eac = 1$ which also contradicts theoretical expectations.
Both of these findings indicate that our understanding of the role of density in microwave photoresistance remains limited calling for further investigations.

%\section{experimental details}
Our 2DEG resides in a 30-nm GaAs/AlGaAs quantum well located about 200 nm below the sample surface.
The structure is doped in a 2-nm GaAs quantum well at a setback of 63 nm on a top side.
The \emph{in situ} back gate consists of an $n^+$ GaAs layer situated 850 nm below the bottom of the quantum well.
Ohmic contacts were fabricated at the corners and midsides of the lithographically defined $1\times1$ mm$^2$ van der Pauw mesa.
The density of the 2DEG was varied from $\ne \approx \na$ to $\nc \times 10^{11}$ cm$^{-2}$.
Over this density range, the low-temperature electron mobility increased from $\mu \approx 0.37$ to $\mu \approx 1.1\times 10^7$ cm$^2$V$^{-1}$s$^{-1}$ \citep{note:mu}, roughly following $\mu \propto \ne^\alpha$, with $\alpha \approx 1.5$ \citep{note:sammon,sammon:2018}.
Microwave radiation, generated by a synthesized sweeper, was delivered to the sample immersed in liquid $^{3}$He via a rectangular (WR-28) stainless steel waveguide. 
The resistance $R$ was measured using a standard low-frequency (a few Hz) lock-in technique.

%FIGURE 1
%%%%%%%%%%%%%%%%%%%%%%%%%
\begin{figure}[t]
%\resizebox{0.5\textwidth}{!}{
\includegraphics{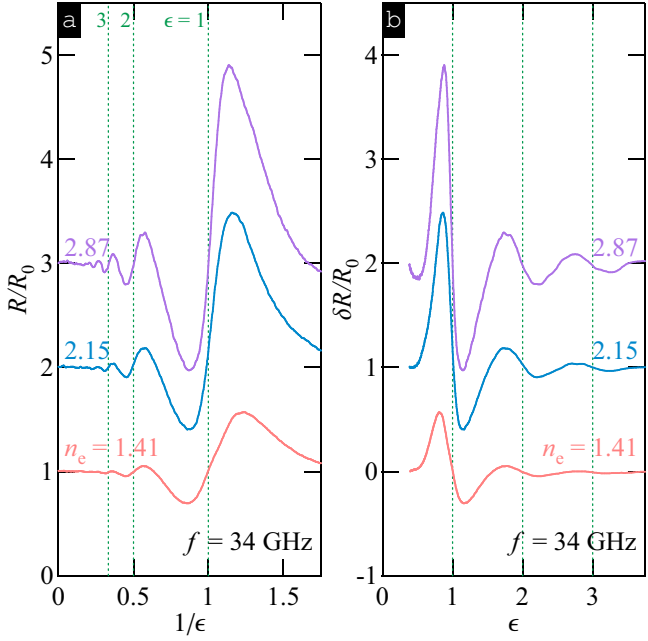}
%}
\vspace{-0.1 in}
\caption{(Color online)
(a) $R/R_0$ vs. $1/\eac$  \citep{note:mass1} at density $\ne \approx \na$ (bottom trace), $\nb$ (middle trace), and $\nc\times 10^{11}$ cm$^{-2}$ (top trace)  at $T = 1.5$ K under irradiation by microwaves of $f = 34$ GHz.
Vertical lines are drawn at $\eac=1,2,3$, as marked.
(b) $\delta R/R_0$ vs $\eac$ under the same conditions as in panel (a).
The traces are vertically offset for clarity by 1.
}
\vspace{-0.1 in}
\label{fig1}
\end{figure}
%%%%%%%%%%%%%%%%%%%%%%%%%

In \rfig{fig1}(a) we present $R/R_0$  vs $1/\eac$ for three different densities, $\ne \approx \na$ (bottom trace), $\nb$ (middle trace), and $\nc\times 10^{11}$ cm$^{-2}$ (top trace), measured at $T = 1.5$ K under irradiation by microwaves of $f = 34$ GHz.
The data show that MIROs become significantly stronger with increasing density.
We next extract the oscillatory correction $\delta R/R_0$ and present the result in \rfig{fig1}(b) as a function of $\eac$.
All three data sets reveal an expected periodicity with $\eac$, in agreement with \req{eq.miro}.

According to \req{eq.miro}, the MIRO amplitude is proportional to the effective microwave power, which was shown to be \citep{chiu:1976,khodas:2008}
\be
\pc(\eac)= \frac {\pc^{0}} 2 \sum\limits_\pm\frac{1}{(1 \pm \eac^{-1})^2+\betao^2},~\pc^{0}=\frac{e^2\Eac^2v_F^2}{\epseff \hbar^2 \omega^4}\,.
\label{eq.pc}
\ee
Here, $\betao\equiv(\omega\tem)^{-1}+(\omega\ttr)^{-1}$, $\ttr = (\m/e)\mu$ is the momentum relaxation time, $\tem^{-1}=\ne e^2/2\sqrt{\epseff}\epsilon_0\m c$ \citep{chiu:1976} is the radiative decay rate, $2\sqrt{\epseff}=\sqrt{\varepsilon}+1$ defines the effective dielectric constant $\epseff$, $\varepsilon=12.8$ is the dielectric constant of GaAs, $v_F$ is the Fermi velocity, and $\Eac$ is the microwave electric field.
The density dependence of $\betao$ has been recently verified in time-resolved measurements of the cyclotron resonance \citep{zhang:2014}.
Within the density range studied in our experiment, $\tau \gg \tem$ and $\betao \approx (\omega\tem)^{-1} \propto \ne$.
However, $\betao$ remains much smaller than unity and, as a result, $\pc$ increases with $\ne$ for all $\eac$ except in close proximity to $\eac = 1$.
We will see, however, that the anticipated increase in $\pc$ is rather small and contributes little to the growth of MIRO shown in \rfig{fig1}. 

%FIGURE 3
%%%%%%%%%%%%%%%%%%%%%%%%%
\begin{figure}[t]
%\resizebox{0.5\textwidth}{!}{
\includegraphics{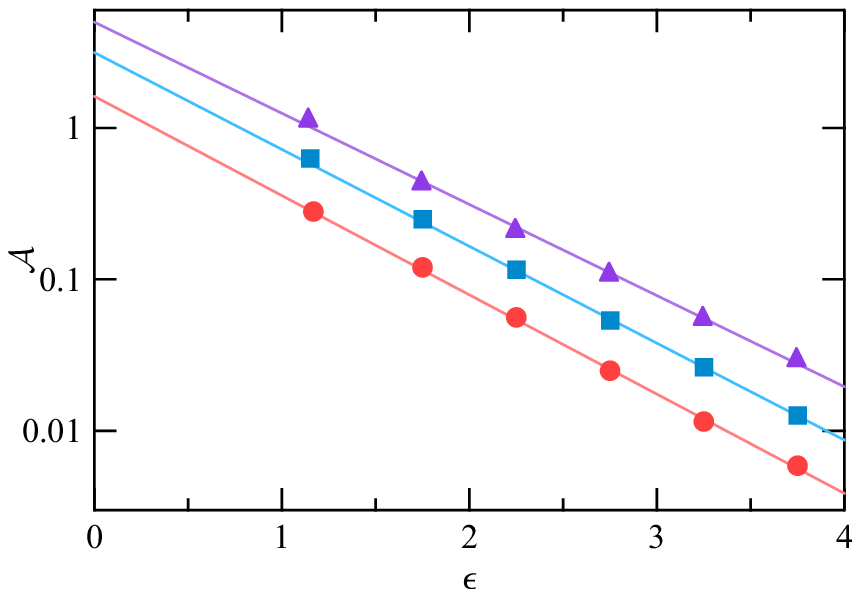}
%}
\vspace{-0.1 in}
\caption{(Color online)
$\A = |\delta R|_{\max}/2\pi\eac \pc R_0$ \citep{note:mass2} vs $\eac$ for $\ne \approx \na$ (circles), $\nb$ (squares), and $\nc\times 10^{11}$ cm$^{-2}$ (triangles) measured at $T = 1.5$ and $f = 34$ GHz.
Fits to the data with $\A_0\exp(-\eac/f\tq)$ (solid lines) yield $\tq \approx \tqa$, $\tqb$, and $\tqc$ ps respectively. 
}
\vspace{-0.1 in}
\label{fig2}
\end{figure}
%%%%%%%%%%%%%%%%%%%%%%%%%
The growth of MIRO with $\ne$ observed in \rfig{fig1} can, in principle, stem from $\tq$ (entering $\lambda$) or $\eta$.
Both of these parameters are readily available from the Dingle analysis.
Following \req{eq.miro}, we introduce a reduced MIRO amplitude $\A = |\delta R|_{\max}/2\pi\eac\pc R_0$ \citep{note:mass2}, where $|\delta R|_{\max}$ is the MIRO amplitude, and present it in \rfig{fig2} as a function of $\eac$ for $\ne = \na$ (circles), $\nb$ (squares), and $\nc\times 10^{11}$ cm$^{-2}$ (triangles).
Fits to the data with $\A_0\exp(-\eac/f\tq)$ (solid lines) yield $\tq \approx \tqa$, $\tqb$, and $\tqc$ ps, respectively, indicating a slight increase of $\tq$ with $\ne$.
In contrast, the intercept of the Dingle plots, $\A_0$, grows substantially with $\ne$. 
As we show below, theory predicts that under our experimental conditions $\A_0$ can only decrease with $\ne$.

\begin{figure}[t]
%\resizebox{0.5\textwidth}{!}{
\includegraphics{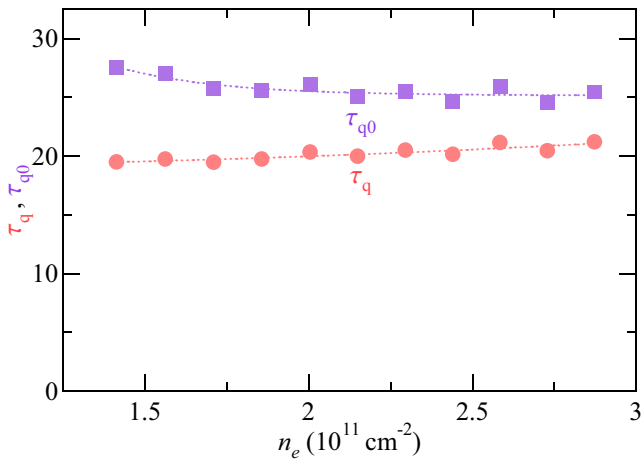}
%}
\vspace{-0.1 in}
\caption{(Color online) 
$\tq$ (circles), obtained from the fits (cf. \rfig{fig2}), and $\tqim$ (squares), calculated using Eqs. \req{eq.tq}and \req{eq.ee}, vs $\ne$.
The dotted lines are guides to the eye.
}
\vspace{-0.1 in}
\label{fig3}
\end{figure}

After repeating the Dingle analysis for other $\ne$, we present the density dependence of $\tq$ (circles) in \rfig{fig3}.
A slight increase of $\tq$ with $\ne$ appears to contradict a recent study \citep{qian:2017b}, which has found a saturation of $\tq$ at $\ne \approx 2 \times 10^{11}$ cm$^{-2}$ and a monotonic decrease at higher $\ne$.
This discrepancy can be alleviated by recalling that Shubnikov$-$de Haas oscillations employed in \rref{qian:2017b} yield only impurity contributions to the quantum lifetime $\tqim$ \citep{martin:2003,adamov:2006}.
The quantum lifetime obtained from MIROs, on the other hand, is reduced by electron-electron scattering.
More specifically \citep{chaplik:1971,giuliani:1982,ryzhii:2003d,ryzhii:2004,hatke:2009a,dmitriev:2009b},
\be
\frac 1 \tq = \frac 1 \tqim + \frac 1 \tee\,.
\label{eq.tq}
\ee
Under the conditions of our experiment the electron-electron scattering rate is given by \citep{chaplik:1971,giuliani:1982,dmitriev:2005} 
\be
\frac\hbar\tee = \frac {\pi k_B^2 T^2} {4 E_F} \ln {\frac  {2 \hbar v_F/a_B}{\pi k_B T}}\,,
\label{eq.ee}
\ee
where $E_F$ is the Fermi energy and $a_B \approx 11$ nm is the Bohr radius in GaAs.
Using Eqs. (\ref{eq.tq}) and (\ref{eq.ee}) we compute $\tqim$ and present the results (squares) in \rfig{fig3}.
The results show that the impurity-limited quantum lifetime decreases slightly with $\ne$, in general agreement with \rref{qian:2017b}.
We note, however, that in our experiment most of this decrease takes place at densities below $\approx 2\times 10^{11}$ cm$^{-2}$.

%%%%%%%%%%%%%%%%%%%%%%%%%
%%%%%%%%%%%%%%%%%%%%%%%%%

\begin{figure}[t]
%\resizebox{0.5\textwidth}{!}{
\includegraphics{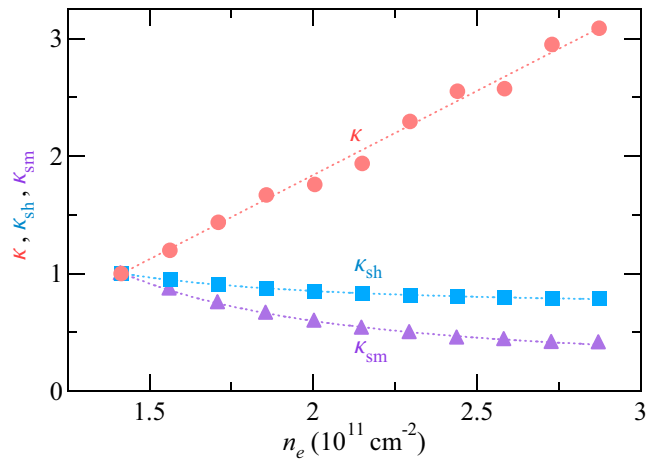}
%}
\vspace{-0.1 in}
\caption{(Color online) 
$\kex$ (circles), obtained from the fits (cf. \rfig{fig2}), $\ksh$ (squares), and $\ksm$ (triangles), calculated using \req{eq.eta} in the sharp and smooth disorder limit, respectively, vs $\ne$.
The dotted lines are guides to the eye.
}
\vspace{-0.1 in}
\label{fig4}
\end{figure}

As already mentioned, \rfig{fig2} also reveals a significant increase of the intercept of the Dingle fits, given by $\A_0$, with increasing $\ne$. 
Since $\A_0 \propto \eta$ \citep{note:a}, this increase reflects the increase in $\eta$, provided that the density dependence of $\pc$ is accurately described by \req{eq.pc}.
To quantify this increase we introduce a parameter $\kex = \A_0(\ne)/\A_0(n_l)$, where $n_l = \na \times 10^{11}$ cm$^{-2}$ is the lowest density studied.
As shown in \rfig{fig4}, $\kex$ (circles) increases by a factor of about 3 over the investigated density range.
This finding is unexpected since,  as we show next, one should anticipate a \emph{decrease} of $\eta$ with increasing $\ne$.

The dimensionless scattering rate $\eta$ is given by \citep{dmitriev:2009b}
\be
\eta = \frac {\ttr}{2\tst} + \frac {2\tin}{\ttr}\,,
\label{eq.eta}
\ee
where the first (second) term represents displacement \citep{durst:2003,lei:2003,vavilov:2004} (inelastic \citep{dmitriev:2005,dmitriev:2009b}) contribution.
Here, $\tin \approx 0.82 \tee$ \citep{dmitriev:2005} and $\ttr/2\tst$ \citep{note:tst} can vary between $\ttr/2\tst = 6(\ttr/\tqim + 3)^{-1}$ (smooth disorder limit) and $\ttr/2\tst = 3/2$ (sharp disorder limit) according to the mixed-disorder model \citep{vavilov:2007,dmitriev:2009b}.
As a result, the relative change in $\eta$ (or $\A_0$) with $\ne$ is expected to fall between $\ksm$ and $\ksh$, given by $\eta(\ne)/\eta(n_l)$ evaluated in the smooth and the sharp disorder limit, respectively.
On a qualitative level, the decrease of $\ttr/2\tst$ with $\ne$ can be expected whenever $\tq/\ttr \ll 1$, i.e., when small angle scattering dominates, which is the case for all modern high-mobility GaAs quantum wells.
This decrease should occur because $\tst^{-1}$ \citep{note:tst} is less sensitive to small-angle scattering than $\ttr^{-1}$ and because the characteristic scattering angle decreases with density.

As shown in \rfig{fig4}, both $\ksh$ (squares) and $\ksm$ (triangles) monotonically decrease with $\ne$.
The decrease in $\ksh$ with $\ne$ occurs solely due to the weakening of the inelastic contribution, given by the second term in \req{eq.eta}.
This weakening, in turn, owes to a superlinear increase in the momentum relaxation time $\tau$, which wins over the slightly sublinear increase in $\tin$  [see \req{eq.ee}]. 
In the smooth disorder limit, characterized by $\ksm$, the decrease becomes larger due to the growing ratio of $\ttr/\tqim$ which enters the denominator of the displacement contribution [first term in \req{eq.eta}].
We thus conclude that regardless of the exact disorder characteristics, theoretical predictions are in contrast with the experimentally obtained $\kex$ (circles) which shows a significant \emph{increase} over the density range studied \citep{note:belt}. 
Our findings were confirmed by measurements using $f = 39.5$ GHz in another sample which are discussed in Supplemental Material \citep{note:sm}.

%%%%%%%%%%%%%%%%%%%%%%%%%
%%%%%%%%%%%%%%%%%%%%%%%%%

\begin{figure}[t]
%\resizebox{0.5\textwidth}{!}{
\includegraphics{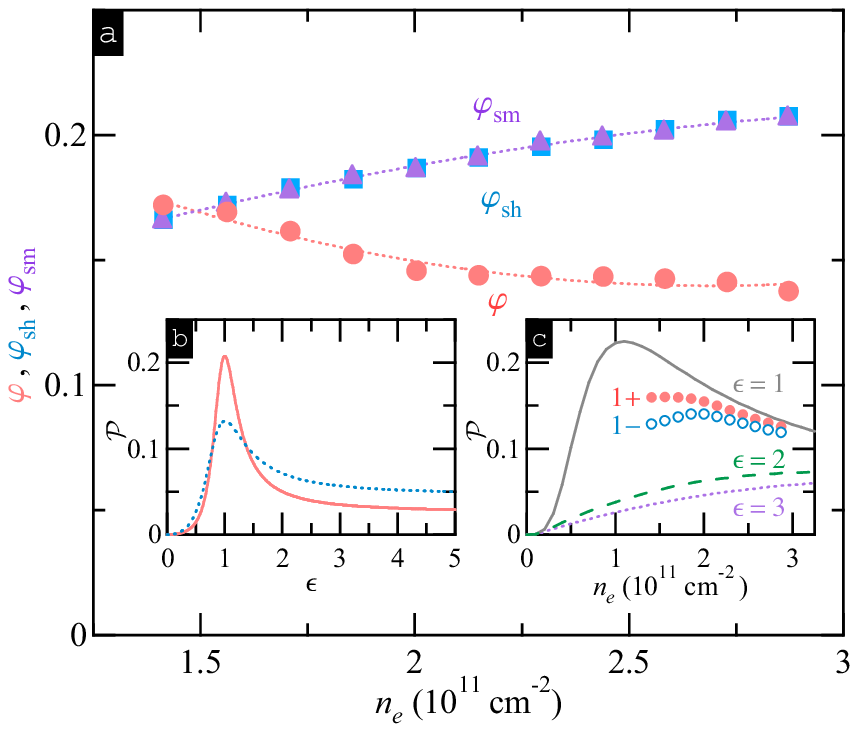}
%}
\vspace{-0.1 in}
\caption{(Color online) 
(a) $\pex$ (circles), $\psh$ (squares), and $\psm$ (triangles) vs $\ne$.
The dotted lines are guides to the eye.
(b) $\pc$ \citep{note:mass2} vs $\eac$ for $\ne = \na \times 10^{11}$ cm$^{-2}$ (solid line) and $\ne = \nc \times 10^{11}$ cm$^{-2}$ (dotted line).
(c) $\pc$ \citep{note:mass2} vs $\ne$ for $\eac = 1$ (solid line), $\eac = 2$ (dashed line), and $\eac = 3$ (dotted line).
Also shown is $\pc$ vs $\ne$ at the first MIRO maximum ($1+$, solid circles) and minimum ($1-$, open circles).
}
\vspace{-0.1 in}
\label{fig5}
\end{figure}

We next examine the effect of density on the positions of the MIRO extrema near the cyclotron resonance.
As noted from the data in \rfig{fig1}(b), these extrema move closer towards $\eac = 1$ with increasing density.
To examine this behavior quantitatively, we introduce a parameter $\pex = (\eac^- - \eac^+)/2$, where $\eac^-$ ($\eac^+$) is the position of the fundamental minimum (maximum).
We then present obtained $\pex$ (circles) in \rfig{fig5}(a) as a function of $\ne$ and observe that it  monotonically \emph{decreases} with $\ne$.
Similar to $\tqim$, the decrease is more pronounced at lower densities.
Theory, however, predicts just the opposite behavior; as illustrated in \rfig{fig5}(a), the calculated values of $\psh$ (squares) and $\psm$ (triangles), representing sharp and smooth disorder limits, respectively, both increase with $\ne$.  
The expected growth of $\psh \approx \psm$ with $\ne$ occurs, for the most part, due to the increase in $\betao$, dominated by $\tem^{-1} \propto \ne$, which controls the sharpness of $\pc$ near $\eac = 1$.
Indeed, as shown in \rfig{fig5}(b), $\pc(\eac)$ is considerably sharper at $\ne = \na \times 10^{11}$ cm$^{-2}$ (solid line) than at $\ne = \nc \times 10^{11}$ cm$^{-2}$ (dotted line).

%%%%%%%%%%%%%%%%%%%%%%%%%
%%%%%%%%%%%%%%%%%%%%%%%%%

It is known that the phase reduction can occur with increasing $\pc$ due to contributions from multiphoton processes \citep{hatke:2011e,shi:2017a}.
This scenario, however, can be ruled out since $\pc$, in fact, decreases with $\ne$ near $\eac = 1$ within the investigated density range.
As shown in \rfig{fig5}(c), $\pc$ at the fundamental MIRO extrema (solid and open circles) exhibits a slight overall decrease within the studied density range, similar to $\pc$ at $\eac = 1$ (solid line).
At higher MIRO orders, $\pc$ monotonically increases, as illustrated by dashed and dotted lines computed for $\eac = 2$ and $\eac = 3$, respectively.
This increase in $\pc$ occurs because $\pc_0 \propto \ne$ while $\betao$ remains relatively small within the studied density range.

%discussion
One somewhat uncertain parameter is $\epseff$ which affects $\betao$ entering $\pc$ given by \req{eq.pc}.
Indeed, the expression we have used is generally valid only when the overall sample thickness greatly exceeds the radiation wavelength, a condition which is not satisfied for the microwave frequency used in our experiment. 
According to \rref{dmitriev:2012}, a better approximation would be using $\epseff = 1$ which would increase the value of $\betao$ by approximately a factor of 2. 
However, any increase in $\betao$ would only weaken (or even reverse) the density dependence of $\pc$ and further increase the disagreement between theory and experiment, both in $\eta$ and $\varphi$. 

Finally, we note that by applying the gate voltage we are not only changing the carrier density but also modifying the confinement potential.
Numerical simulations show that the 2DEG is pulled away from the top interface towards the center of the quantum well and becomes wider with increasing density.
Whether or not such a change of the confinement plays any significant role in the observed enhancement of the MIRO amplitude is unclear at this point and is left for future studies.
To investigate this possibility, it would be interesting to perform measurements in different structures, such as heterojunction-insulated gate field-effect transistors, in which confinement becomes stronger with increasing carrier density.

%summary
In summary, we have investigated the effect of the carrier density $\ne$ on the MIRO amplitude in a high-mobility modulation-doped GaAs/AlGaAs quantum well equipped with an \emph{in situ} back gate.
Our main finding is a significant \emph{growth} of the MIRO amplitude with increasing density.
A Dingle analysis shows that this increase originates primarily from $\pc\eta$ entering \req{eq.miro} and not from a slight increase of $\tq$.
This finding is in conflict with theoretical expectations which predict a modest increase of $\pc$ and a \emph{decrease} of $\eta$ with increasing density.
We further find that the MIRO extrema near the cyclotron resonance move toward each other with increasing $\ne$ whereas the theory predicts just the opposite behavior.
These findings indicate that our understanding of microwave photoresistance is still lacking and needs further examination.

\begin{acknowledgements}
We thank Q. Shi and Q. Ebner for assistance with experiments and M. Sammon for discussions.
The work at Minnesota was supported by the NSF Award No. DMR-1309578 (experiments discussed in the main text) and by the U.S. Department of Energy, Office of Science, Basic Energy Sciences, under Award No. ER 46640-SC0002567 (complementary experiments discussed in Supplemental Material).
The work at Purdue was supported by the U.S. Department of Energy, Office of Science, Basic Energy Sciences, under Award No. DE-SC0006671.
\end{acknowledgements}

\small{$^\#$Present address: Microsoft Station-Q at Delft University of Technology, 2600 GA Delft, The Netherlands}


\begin{thebibliography}{54}
\bibitem[{not({\natexlab{a}})}]{note:mu}
\bibinfo{note}{Mobility values were estimated using the standard van der Pauw
  method. The zero-field resistance anisotropy ratio in our samples was found to be
  less than 2.}
\bibitem[{not({\natexlab{d}})}]{note:mass2}
\bibinfo{note}{$\pc$ was calculated using \req{eq.pc} with $\Eac = 1$ V/cm and
  $\m = 0.067m_0$ entering $\eac$ and $\tem$.}
\bibitem[{not({\natexlab{e}})}]{note:a}
\bibinfo{note}{Microwave power is not exactly known and $\A_0$ differs from
  $\eta$ by a numerical factor.}
\end{thebibliography}
\end{document}

\title{Supplemental Material to ``Effect of density on microwave-induced resistance oscillations in back-gated GaAs quantum wells''}

\author{X. Fu}
\affiliation{School of Physics and Astronomy, University of Minnesota, Minneapolis, Minnesota 55455, USA}
\author{M.~D.~Borisov}
\affiliation{School of Physics and Astronomy, University of Minnesota, Minneapolis, Minnesota 55455, USA}
\author{M.~A.~Zudov}
\affiliation{School of Physics and Astronomy, University of Minnesota, Minneapolis, Minnesota 55455, USA}
\author{Q.~Qian}
\affiliation{Department of Physics and Astronomy and Birck Nanotechnology Center, Purdue University, West Lafayette, Indiana 47907, USA}
\author{J.\,D. Watson}
\affiliation{Department of Physics and Astronomy and Birck Nanotechnology Center, Purdue University, West Lafayette, Indiana 47907, USA}
\author{M.\,J. Manfra}
\affiliation{Department of Physics and Astronomy and Birck Nanotechnology Center, Purdue University, West Lafayette, Indiana 47907, USA}
\affiliation{Station Q Purdue, Purdue University, West Lafayette, Indiana 47907, USA}
\affiliation{School of Materials Engineering and School of Electrical and Computer Engineering, Purdue University, West Lafayette, Indiana 47907, USA}
%\received{\today}

\begin{abstract}
This Supplemental Material presents additional data obtained in a different sample using microwave frequency of $f = 39.5$ GHz which further support conclusions presented in the main text.
\end{abstract}
\received{\today}
%\pacs{73.40.-c, 73.21.-b, 73.43.-f}
\maketitle

%\emph{Experimental details.}
To confirm our findings we have investigated another sample fabricated from a 30-nm GaAs/AlGaAs quantum well equipped with an \emph{in situ} back gate situated 850 nm below the bottom of the quantum well.
Ohmic contacts were fabricated at the corners and midsides of the lithographically-defined $1\times1$ mm$^2$ Van der Pauw mesa.
The density of the 2DEG was varied from $\ne \approx 1.6$ to $2.4 \times 10^{11}$ cm$^{-2}$.
Unfortunately, neither higher nor lower densities were accessible in this device because of the gate leakage.
Microwave radiation of frequency $f = 39.5$ GHz was delivered to the sample immersed in liquid $^{3}$He via a rectangular (WR-28) stainless steel waveguide. 
The resistance $R$ was measured at temperature $T = 1.6$ K using a standard low-frequency (a few Hz) lock-in technique.

\begin{figure}[b]
\includegraphics{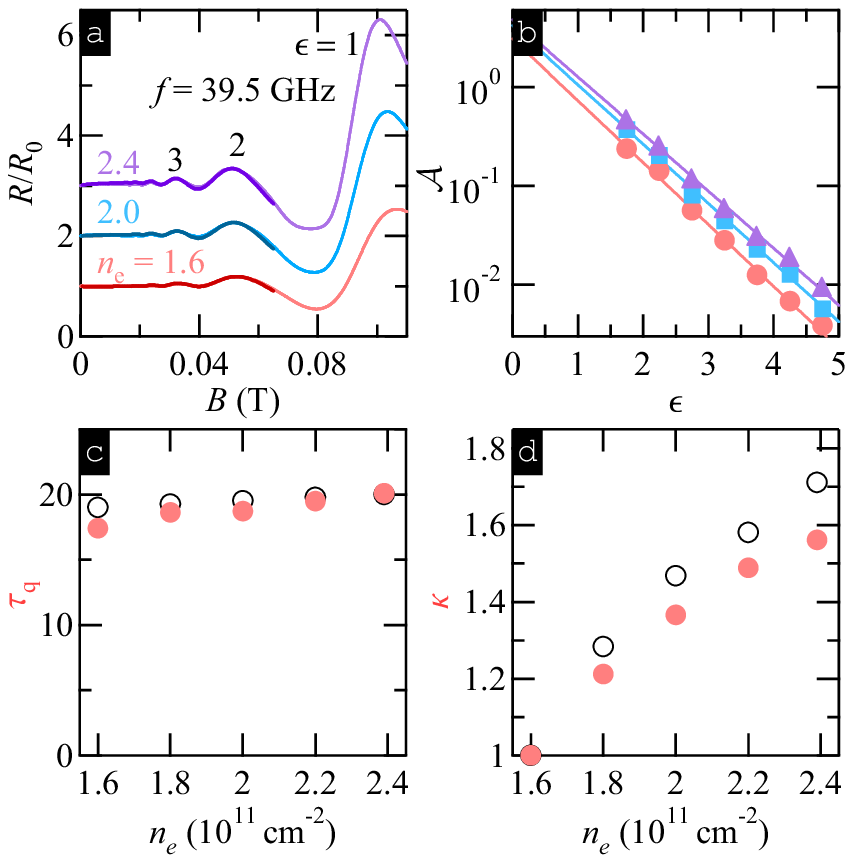}
\vspace{-0.15 in}
\caption{(Color online)
(a) $R/R_0$ vs. $B$  at density $\ne \approx 1.6$ (bottom trace), $2.0$ (middle trace), and $2.4\times 10^{11}$ cm$^{-2}$ (top trace) under irradiation by microwaves of $f = 39.5$ GHz.
Traces are vertically offset by 1 for clarity.
Solid lines are fits to Eq. (1). 
(b) $\A = |\delta R|_{\max}/2\pi\eac \pc R_0$ vs $\eac$ for $\ne \approx 1.6$ (circles), $2.0$ (squares), and $2.4 \times 10^{11}$ cm$^{-2}$ (triangles) extracted from the data in panel (a).
Lines are fits to the data with $\A_0\exp(-\eac/f\tq)$. 
(c) $\tq$, obtained from the fits in panel (b) (solid circles) and from the fits in panel (a) (open circles) vs $\ne$.
(d) $\kappa = \A_0(\ne)/\A_0(n_l)$, with $n_l = 1.6 \times 10^{11}$ cm$^{-2}$, obtained from the fits in panel (b) (solid circles) and from the fits in panel (a) (open circles) vs $\ne$.
}
\vspace{-0.15 in}
\label{fig1}
\end{figure}

%\emph{Results.}
In \rfig{fig1}(a) we present $R/R_0$  vs $B$ for three different densities, $\ne \approx 1.4$ (bottom trace), $2.0$ (middle trace), and $2.4\times 10^{11}$ cm$^{-2}$ (top trace), and observe that MIRO become stronger with $\ne$.
In \rfig{fig1}(b) we show a reduced MIRO amplitude $\A = |\delta R|_{\max}/2\pi\eac\pc R_0$, where $|\delta R|_{\max}$ is the MIRO amplitude, as a function of $\eac = 2\pi f/\oc$ for $\ne = 1.6$ (circles), $2.0$ (squares), and $2.4\times 10^{11}$ cm$^{-2}$ (triangles).
Fits to the data with $\A_0\exp(-\eac/f\tq)$ (solid lines) yield $\tq$ and $\A_0$.
As shown in \rfig{fig1}(c), extracted $\tq$ (filled circles) increases only slightly with $\ne$.
In contrast, as illustrated in \rfig{fig1}(d), $\kappa = \A_0/\A_0(n_l)$ (filled circles), where $n_l = 1.6 \times 10^{11}$ cm$^{-2}$, grows substantially with $\ne$. 
To test the reliability of the fitting procedure, we also directly fit the low field part of the data, $R/R_0$ vs $B$ shown in \rfig{fig1}(a), with Eq. (1) (see main text). 
The fits, shown by thick lines in \rfig{fig1}(a), produce a slightly higher $\tq$ [cf. open circles in \rfig{fig1}(c)] and a somewhat larger increase in $\kappa$ [cf. open circles in \rfig{fig1}(d)].
Nevertheless, both data reduction methods reveal growth of $\kappa$ with $\ne$ which is similar in magnitude to that presented in the main text within the same density range. 
\newpage

%%%%%%%%%%%%%%%%%%%%%%%%%%%%%%%%%%%%%%%%%%%%
%%%%%%%%%%%%%%%%%%%%%%%%%%%%%%%%%

%\bibliography{../../bibRMP1qs.1_2,footnotes2}
%\bibliographystyle{../../apsrev}
%\bibliographystyle{apsrevmod}
%\bibliography{bibRMP1qs.1_2}

\vspace{-0.3 in}